\documentclass[aps,prd,preprint,superscriptaddress]{revtex4-1}
\usepackage{graphicx,array,dcolumn}
\usepackage{array}
\usepackage{amsmath,setspace,latexsym}
\usepackage[colorlinks=true
  ,urlcolor=blue
  ,anchorcolor=blue
  ,citecolor=blue
  ,filecolor=blue
  ,linkcolor=blue
  ,menucolor=blue
  ,pagecolor=blue
  ,linktocpage=true
  ,pdfproducer=medialab
  ,pdfa=true]{hyperref}
\usepackage{physics}
\usepackage{xcolor}
\usepackage{outlines}
\usepackage{booktabs}
\usepackage{float}
\usepackage{caption}
\usepackage{subcaption}
\usepackage{bbold}
\usepackage{multirow}
\usepackage{booktabs}
\usepackage{textcomp}
\usepackage{placeins}
\usepackage{multirow}
\usepackage[export]{adjustbox}

\newlength{\figurewidth}
\newenvironment{equations}{\equation\aligned}{\endaligned\endequation}  
\newcommand{\beq}{\begin{equations}}
\newcommand{\eeq}{\end{equations}}
\newcommand{\bc}{\begin{cases}}
\newcommand{\ec}{\end{cases}}
\newcommand{\boe}{\begin{outline}[enumerate]}
\newcommand{\eo}{\end{outline}}
\def\l{\left(}
\def\r{\right)}
\def\ll{\left[}
\def\rr{\right]}

\def\la{\mathcal{L}}

\def\={&=}
\def\p{\partial}
\def\g{g_{\mu\nu}}

\def\tg{\textcolor{green}}

\def\tm{\textcolor{magenta}}

\def\-{\item[-]}

\def\p{\partial}

\def\uk{u_{k}}
\def\rk{R_{k}}
\def\rkk{\mathbb{R}_{k}}
\def\tlam{\tilde{\lambda}^{2}}
\def\tu{\tilde{u}_{k}}
\def\tphi{\tilde{\phi}}
\def\ga2{\Gamma^{(2)}}
\def\lam2{\lambda^{2}}
\def\tm{\tilde{m}^{2}}
\def\tg{\tilde{g}}
\def\tfg{\tilde{g}^{*}}

\def\gak{\Gamma_{k}}
\def\ggk{\mathbb{\Gamma}_{k}}
\def\gg2{\mathbb{\Gamma}^{(2)}_{k}}
\def\ggg{\mathbb{\Gamma}^{(3)}_{k}}

\def\weq{Wetterich equation (Eq. \ref{weq}) }
\def\z1{Z_{1}}

\def\g3{\Gamma^{(3)}}
\def\trh{\tilde{\rho}}
\def\tz{\tilde{Z_{1}}}
\def\tj{\tilde{J}_{l}}
\def\tk{\tilde{\kappa}}
\def\zz{Z_{2}}

\def\tjnl{\tilde{J}_{nl}}
\def\deff{D_{\text{eff}}}

\makeatother

\begin{document}
\onehalfspacing
%
\title{
RG studies of scalar-field models of long-range interactions
}
\setlength{\figurewidth}{\columnwidth}
 \author{Alfio M. Bonanno}
\email{alfio.bonanno@inaf.it}
\affiliation{INAF, Osservatorio Astrofisico di Catania, via S. Sofia 78, 95123 Catania, Italy}
\affiliation{INFN, Sezione di Catania, via S. Sofia 64, 95123,Catania, Italy}

\author{S. R. Haridev}
\email{haridevsr@iisc.ac.in}
\affiliation{Center for High Energy Physics, Indian Institute of Science, C V Raman Road, Bangalore 560012, India}
\affiliation{Department of Physics, Indian Institute of Science, C V Raman Road, Bangalore 560012, India}

\author{Gaurav Narain}
\email{gnarain@iisc.ac.in}
\affiliation{Center for High Energy Physics, Indian Institute of Science, C V Raman Road, Bangalore 560012, India}

\begin{abstract}

In this work we studies the long-range interactions in non-gravitational field theories and their behaviour in the deep infrared. To model such effects, we consider a nonlocal scalar theory obtained by adding a $\phi\Box^{-1}\phi$ term to the local action. Using the functional renormalisation group, we analyse its infrared fixed-point structure. Within the LPA, we show that nonlocality modifies phase-transition patterns and can induce symmetry breaking. Extending the LPA beyond polynomial truncations, we examine the convexity property of the effective potential as $k\rightarrow 0$ and find that the flow becomes singular for $\lambda^{2}>0$ before reaching the deep infrared. In the LPA$'$ framework, we find that the infrared-stable fixed point is the nonlocal Gaussian fixed point. We then generalise the model to $\phi\Box^{\sigma/2}\phi$ and analyse how the infrared properties depend on $\sigma$. With appropriate scaling choices, we show that the infrared behaviour remains unchanged up to $\sigma=d/2$ and follows Sak’s prediction up to $\sigma=2$. Finally, we study higher-derivative cases within the LPA, focusing on $\sigma=4$, which corresponds to isotropic Lifshitz criticality, and obtain results consistent with earlier work.
\end{abstract}

\vspace{5mm}


\maketitle

\tableofcontents

\section{Introduction}
Locality is widely considered as a fundamental principle of nature. However, quantum corrections can naturally introduce nonlocal interactions in the effective description. Such nonlocalities often emerge when massless or light degrees of freedom are integrated out of an underlying local theory. Moreover, nonlocality arises naturally in quantum gravity frameworks like string theory, where it helps address divergences and provides a mechanism for UV completion.  Introducing nonlocal interactions can significantly alter the fixed-point structure of a theory, reshaping both its low-energy (IR) behaviour and high-energy dynamics by introducing new stable fixed points, changes in scaling laws, or suppression of divergences \cite{RevModPhys.95.035002, Belgacem_2018}.  This has significant implications for cosmological models, where nonlocal corrections to gravity can affect large-scale structure of the universe. Similarly, in condensed matter systems, low-energy behaviour of a model with long-range interactions is effectively equivalent to a nonlocal theory \cite{RevModPhys.95.035002,Defenu_2020}. In general, nonlocal interactions allowed by the symmetries of the theory are equally considerable as local interactions \cite{Niedermaier2006}. These nonlocal interactions give interesting results in the asymptotic safety of quantum gravity, where it dynamically solves the IR problems of quantum gravity \cite{Wetterich1998, PhysRevD.66.125001, sym16081074}. Thus, nonlocal field theories offer a richer framework to explore novel IR phenomena across different areas of physics.

Recent studies have explored the nonlocal aspects in various contexts. Nonlocal theories allow the construction of explicit models where symmetry breaking persists at arbitrarily high temperatures due to long-range interactions \cite{10.21468/SciPostPhys.12.6.181}. These models could advance the realization of high-temperature superconductors and can also improve our understanding of phase transitions in the early universe. The nonlocal modification of gravity is used in cosmological models to explain inflation or late-time cosmic acceleration through infrared nonlocality without the need for additional degrees of freedom \cite{PhysRevD.93.063008,PhysRevD.97.083523,universe4080082,PhysRevLett.99.111301}. Due to their inherent long-range correlations, nonlocal interactions can also significantly affect the entanglement structure of a theory, a topic extensively discussed in literature, including in the context of de Sitter and anti- de Sitter spacetime \cite{Narain_2019,PhysRevD.99.125012}. Furthermore, analyzing infrared behaviour or the low-energy physics of nonlocal theories—particularly the model we will employ—has yielded promising results, notably providing a resolution to the IR problem of the de Sitter propagator \cite{NARAIN2019143}. It is also established that certain classes of nonlocal gravity theories can be constructed to be free of ghosts and singularities. Remarkably, despite their nonlocal nature, these theories recover the correct infrared limit, Einstein’s general relativity in the IR limit \cite{PhysRevLett.108.031101, biswas2013nonlocaltheoriesgravityflat}. Moreover, by incorporating nonlocal terms, one can construct super-renormalizable models that improve the ultraviolet behaviour of gravity while maintaining unitarity \cite{PhysRevD.86.044005, doi:10.1142/S0218271817300208,Moffat2011}. These nonlocal terms typically introduces exponential or similar suppressive factors in the propagators at high energies, effectively softening divergences and improving the UV behaviour of the theory. Additionally, the implications of nonlocality have been explored in the context of the Unruh effect and the black hole information loss paradox, offering novel insights into these foundational problems \cite{KAJURI2019319, PhysRevD.95.101701}. Together, these developments highlight the potential of nonlocal theories to address both infrared and ultraviolet challenges in a unified framework.

In this work, we are trying to explore the fixed point structure of a single scalar field theory with nonlocality introduced through $\phi\Box^{-1}\phi $ interactions. This model serves as a toy model for nonlocal modifications of gravity, similar to the Woodard-Deser model \cite{PhysRevLett.99.111301} and related approaches \cite{PhysRevD.93.063008,sym16081074,NOJIRI2008821}, where understanding the infrared (IR) dynamics is crucial for consistency and its viability as a dark energy alternative. In condensed matter physics, the model can be interpreted as an effective description of systems with strong long-range interactions \cite{RevModPhys.95.035002,Defenu_2020}, and our focus is on exploring its universal behaviour in the low-energy limit and assessing how smoothly it connects to the established long-range behaviour discussed in \cite{PhysRevE.92.052113, Defenu_2020}. The model can be reformulated as a local theory involving two scalar fields. This property allows us to validate the results obtained from the nonlocal formulation by comparing them with those of the well-established local theory. Our objective in this analysis is to understand the complete phase space structure of the theory, focus on low-energy behaviour, and relate the model to known models in the literature. To achieve our goal, we use the standard renormalization group method to understand physics at different energy scales. Given the presence of long-range correlations in the model, it is most effective to employ exact renormalization group methods. In particular, we utilize the functional renormalization group (FRG) approach, which is well-suited for capturing non-perturbative features of the theory \cite{WETTERICH199390, BERGES2002223}. Using the FRG, we are also able to extend the analysis to a more general class of theories with $\phi\Box^{\sigma/2}\phi$ interactions. In general, these theories are not unitary; one may have to do the complete spectral analysis to study their unitary properties. However, the renormalisation analysis of such models remains well motivated where one is interested in the universal scaling properties, operator dimensions, and the structure of fixed points rather than the time evolution of the model \cite{PhysRevD.98.085005, ZAPPALA2017213, BONANNO2015501}. These models also naturally arise as an effective description for long-range interactions and exhibit well-defined critical behaviour \cite{PhysRevE.92.052113,RevModPhys.95.035002}.

This paper is organized as follows. In the first section \ref{model}, we introduce the details of our model and the FRG method employed in our analysis. We begin with the leading approximation in FRG, known as the local potential approximation (LPA), where the nonlocal coupling is treated as an external parameter. Within this framework, we examine the impact of the nonlocal coupling on the phase transition. Subsequently, we extend the analysis by incorporating wave function renormalization (LPA\textquotesingle). The section concludes by demonstrating that the infrared-stable fixed point of the theory is the nonlocal Gaussian fixed point (NL GFP). In the next section \ref{seclocal}, we re-derive the same results using the equivalent local theory formulated with two scalar fields. Then, in section \ref{arbsigma}, we generalize our model to include arbitrary interactions of the form $\phi \Box^{\sigma/2}\phi$, establishing a smooth connection between our framework and existing long-range models in the literature, as well as higher-derivative theories. Finally, in the last section \ref{seccon}, we summarize our findings and discuss potential extensions of this work. In the Appendix \ref{App1}, we derive the flow equations for all parameters in our nonlocal theory, both explicitly within the nonlocal formulation and through the corresponding local theory. In Appendix \ref{appshoot}, we solve the flow equation for the nonlocal model using an asymptotic solution, confirming consistency with the results presented in the main text through spike analysis. Throughout this paper, we work in natural units with $\hbar= c= 1$, and in the Euclidean framework with metric convention $(+,+,+,+)$.
\section{Non-Local Model}\label{model}
In this section, we introduce a model of scalar field theory with non-local interactions. We examine the functional space structure of the model using the functional renormalization group method. The non-local Lagrangian density for the scalar field is
\newpage
\begin{equation}
\label{mainL}
\la = \frac{1}{2}\l\p\phi\r^{2}+\frac{1}{2}m^{2}\phi^{2}+\frac{\lambda^{2}}{2}\phi\frac{1}{-\Box}\phi + U(\phi).
\end{equation}
Here, $\Box$ represents the Laplacian operator in flat space, and $\lambda$ is the real-valued non-local coupling. In momentum space, the non-local terms scale as $p^{-2}$, dominating the infrared limit. The theory contains complex mass poles, as discussed in \cite{Narain_2019, PhysRevD.99.125012, NARAIN2019143},  and have been extensively studied in the literature \cite{10.1143/PTP.44.272,Nagy1970, LEE1969209}. In this work, we are considering the model as an effective description of a more fundamental ultra violet complete theory. One can also describe this non-local theory using a local theory that involves two scalar fields. We will explore this in the section \ref{seclocal}.

To study the functional space structure of the theory, we introduce a scale-dependent effective action, $\gak$, which depends on an infrared cutoff $k$. As $k \rightarrow \infty$, the effective action approaches the classical action. In the limit as $k \rightarrow 0$, it gives the full quantum action. For intermediate scales, $\gak$ provides the mean-field free energy.  The flow equation for the effective average action (EAA) is given by the Wetterich flow equation as \cite{WETTERICH199390}
\beq\label{weq}
\p_{t}\Gamma_{k} \= \frac{1}{2}\tr \l \frac{\p_{t}\rk}{\ga2_{k}+\rk}\r,
\eeq
where the trace is over all the spacetime and internal indices, $t=-\log\l k/k_{0}\r$,  $\rk$ is the regulator and $\ga2_{k}$ is the two point vertex function. The addition of the cutoff function has not modified the vertices, its only effect is to replace the original inverse propagator by the cutoff inverse propagator so that we can obtain a well behaved modified propagator for the theory.  The simplest ansatz for $\gak$ is the Local Potential Approximation (LPA) \cite{BERGES2002223,Wipf2021, doi:10.1142/10369}, which is given as
\beq
\gak = \int_{x}\frac{1}{2}\l\p\phi\r^{2}+\frac{\lambda^{2}}{2}\phi\frac{1}{-\Box}\phi + u\l\rho\r,
\eeq
where $2\rho =\phi^{2} $.  In order to solve the \weq one has to choose the form of $\rk$ to obtain the flow equation for different parameters. In our model we choose the regulator function as
\beq\label{regulator1}
\rk = \ll \l k^{2}-q^{2}\r+\lambda^{2}\l\frac{1}{k^{2}}-\frac{1}{q^{2}}\r\rr\theta\l k^{2}-q^{2}\r.
\eeq
\begin{figure}[h]
\centering
\includegraphics[scale=0.8]{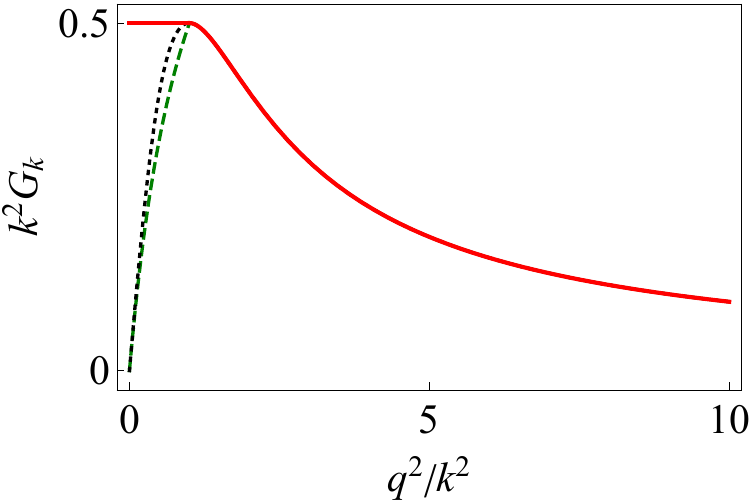}
\caption{Red solid line shows the modified cutoff propagator using Eq. \ref{regulator}, Dashed green line shows the modified propagator with standard Litim cutoff and black dotted line show the original propagator.}
\label{propfig}
\end{figure}
Note that as $q^{2}\rightarrow 0$ $\rk\rightarrow -\infty$; however, the modified propagator is well-behaved, as shown in Fig. \ref{propfig}, which is the necessary condition \cite{doi:10.1142/10369}. Using this regulator function and constant field approximation, one can obtain
\beq
\label{unscalukflow}
\p_{t}\lam2 \=0, \quad
\p_{t}\uk = -\mu_{d}k^{d}\l k^{2}-\frac{\lambda^{2}}{k^{2}}\r\l k^{2}+\frac{\lambda^{2}}{k^{2}}+m^{2}\r^{-1},
\eeq
where $\mu_{d}^{-1}=\l 4\pi\r^{d/2}\Gamma\l d/2+1\r$ and $m^{2}=\uk''\l\phi\r$. In deriving $\p_{t}\lam2 =0$, we only use the analyticity of non local propagator, so the statement is independent of the cutoff function (see Appendix.\ref{App1}).
\vspace{-1em}
\subsection{Nonlocal coupling as external parameter}
\begin{figure}[h!]
    \centering
     \begin{subfigure}{0.5\textwidth}
        \includegraphics[scale=0.38]{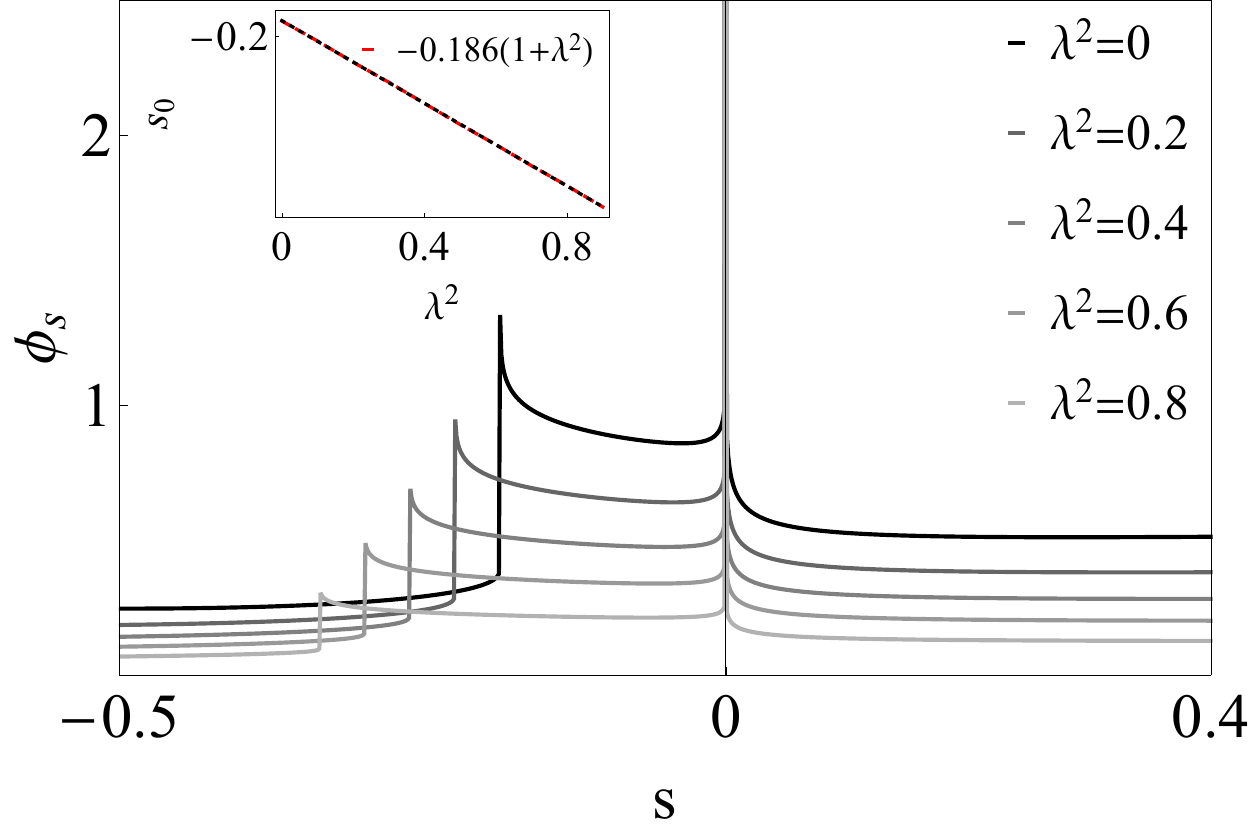} 
        \phantomsubcaption\label{figspike3d}
        \centering (a)
    \end{subfigure}
    \begin{subfigure}{0.48\textwidth}
        \includegraphics[scale=0.32]{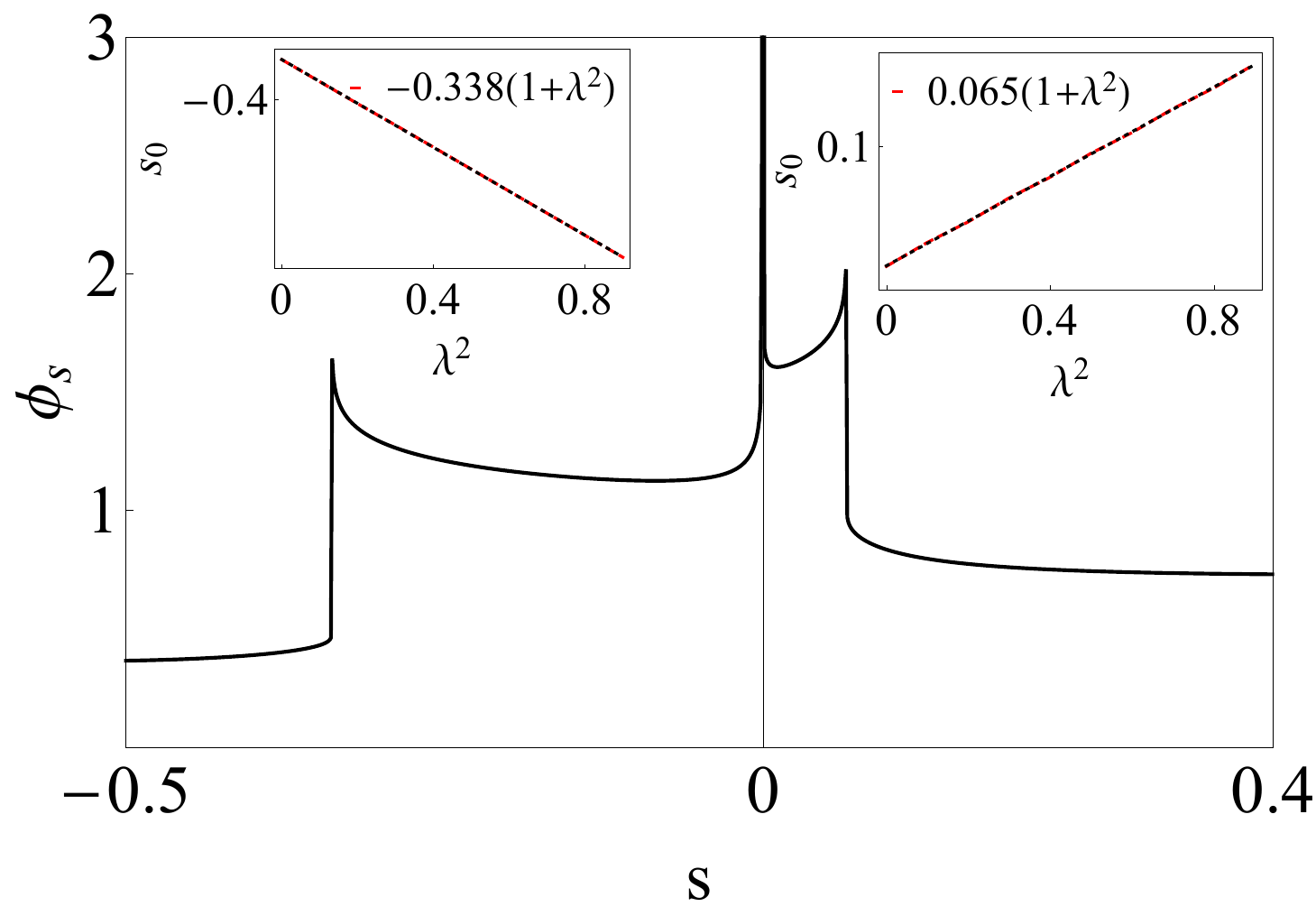}
        \phantomsubcaption\label{figspikelowd}
        \centering (b)
    \end{subfigure}
     \caption{(a) Spike plot for different values of $\tlam$ in $d=3$. The black line corresponds to the local theory $\tlam=0$ and as we increase the $\tlam$ the spike shift towards left. The plot corresponds to $\tlam=0$ to $\tlam=1$ at an interval of $0.2$. In the inset we plot the variation $s_{0}$ corresponds to WF fixed point as a function of non-local coupling ($\tlam$). We also shows the best line fit for the data point, which explicitly shows the linear behavior. (b) Spike analysis and the linear shift of non Gaussian fixed points in $d=2.6$ }
     \label{extspike}
\end{figure}

In this subsection, we consider the non local coupling as an extrenal parameter which is not flowing.  We are interested in the fixed point solutions of the theory, so we can scale the parameters in the theory as $\phi =k^{\frac{d-2}{2}}\tphi$, $\uk=k^{d}\tu$ and $\lam2=k^{4}\tlam$ which leads to scaled flow equation as
\beq\label{scaledflowu}
\p_{t}\tu = d \tu -\frac{d-2}{2}\tphi\tu' - \mu_{d}\l 1-\tlam\r \l1+\tm +\tlam\r^{-1}.
\eeq
Without any further approximation on the form of potential, one can solve the above equation (Eq. \ref{scaledflowu}) using the spike analysis \cite{Codello_2012} for different values of the non local coupling as shown in Fig. \ref{extspike}. Plots in Fig. \ref{figspike3d} corresponds to the three dimensional case, where the spike at non zero value of $s$ corresponds to the Wilson Fischer fixed point. One can see that the spike corresponds to the Wilson Fischer fixed point moves away as we increase the value of non local coupling. The $s$ value corresponds to each spike gives the value of mass parameter ($u''\l \phi=0\r = s_{0}$) at the corresponding fixed points. The inset plot in Fig. \ref{figspike3d} shows the shift in $s_{0}$ as we increase the non local coupling $\tlam$. The linear shift of the non-Gaussian fixed point with increasing non-local coupling holds even in lower dimensions, as illustrated in Fig. \ref{figspikelowd}. 
Now we do the fixed point analysis using the polynomial truncation, where we assume that
\beq\label{u}
\tu(\phi) = \sum_{n=0}^{N}\frac{\tg_{2n}}{(2n)!}\tphi^{2n}.
\eeq
Using polynomial ansatz for the potential (Eq. \ref{u}) in the flow equation for the potential (Eq. \ref{scaledflowu}) we can obtain the flow equation for the local couplings in Eq. \ref{u}. One can solve these flow equation for arbitrary trunction in Eq. \ref{u} to obtain the fixed point values of the local couplings as
\beq\label{fixedpointd}
\tfg_{0}\= \frac{\mu_{d}}{d}\frac{1-\tlam}{\Delta_{*}}, \quad
\tfg_{4} = \frac{2}{\mu_{d}}\frac{\tfg_{2}\Delta_{*}^{2}}{\tlam-1},\quad
\tfg_{6} = \frac{4-d}{\mu_{d}}\frac{\tfg_{4}\Delta_{*}^{2}}{\lambda^{2}-1}+\frac{6\tilde{g}^{*2}_{4}}{\Delta_{*}},...\\
\eeq
where $\Delta_{*}=1+\tlam+\tfg_{2}$. Note that these solutions are true in all order in $N$ and from the spike analysis we know $g_{2}^{*}=s_{0}$ as a function of $\tlam$, which is true for arbitrary truncation. Using the results in Eq. \ref{fixedpointd} one can obtain the fixed point solution of the local coupling which is true for arbitrary truncations. 
\begin{figure}[h!]
\centering
 \includegraphics[scale=0.8]{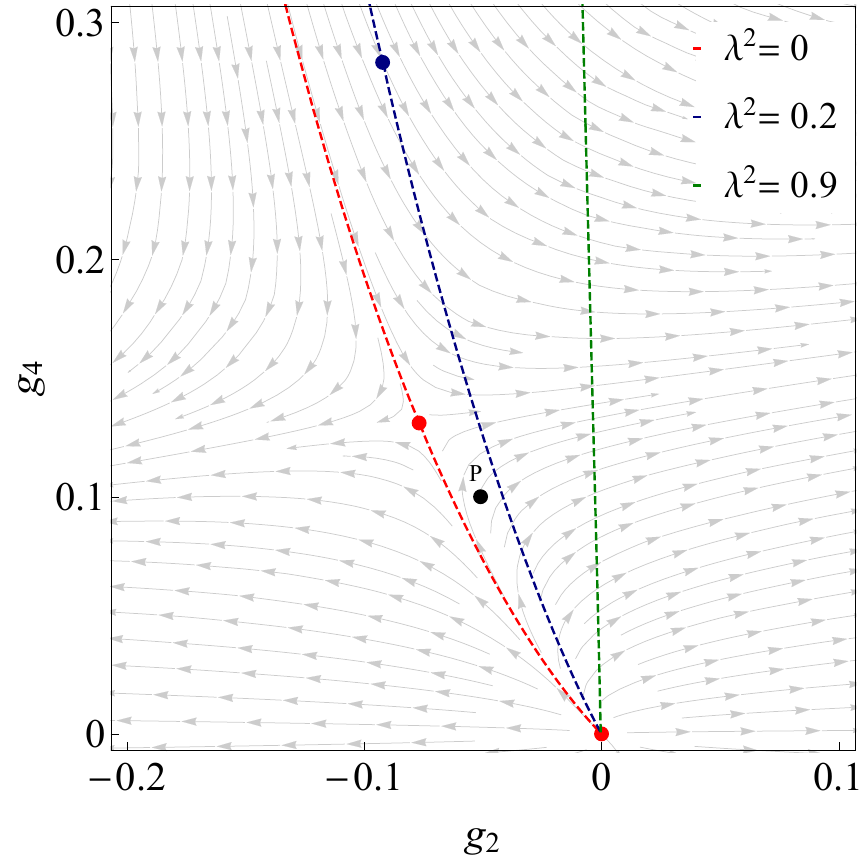} 
  \caption{RG-flow diagram with parameters $g_{2}$ and $g_{4}$ for $N=2$ truncation and $d=3$. Separatrix of different phases for different values of $\lam2$ are shown in different colors. Red dashed lines shows the separatrix for $\lam2=0$, the region flowing towards large positive $g_{2}$ corresponds to the symmetric phase and the other phase corresponds to the broken symmetry phase.  Green and blue dashed lines shows the separatrix for non zero values of $\lam2$.  The red dot corresponds to the Gaussian and WF FP with $\lam2=0$ and the blue dot corresponds to WF FP with $\lam2=0.2$. Point "P" represents a symmetric phase for $\lam2=0$ and broken phase for $\lam2\geq 0.1$}
  \label{extphase}
\end{figure}
Now we calculate the critical exponents of the fixed point, which are the physically measurable quantities. Critical exponets characterize a fixed point and provide insights into the stability of the fixed point.  One can obtain the critical exponets from the eigen values of the stability matrix given as
\beq
S_{ij}\= \frac{\p \beta_{i}}{\p\tg_{j}}\Big|_{*}.
\eeq
\newpage
For the primary analysis we keep truncation upto quartic order in the potential (i.e. $N=2$ in Eq. \ref{u}), which gives the stability matrix as
\beq
S_{ij} = \begin{pmatrix}
-d & \mu_{d}\frac{\lambda^{2}-1}{\Delta_{*}^{2}} & 0\\
0 & -2 -\frac{4\tfg_{2}}{\Delta_{*}} & \mu_{d} \frac{\lambda^{2}-1}{\Delta_{*}^{2}}\\
0 & \frac{72 \tilde{g}_{2}^{*2}}{\mu_{d}\l \lambda^{2}-1\r} & d-4- \frac{24\tfg_{2}}{\Delta_{*}}
\end{pmatrix}.
\eeq
Note that for fixed a dimensions, the stability matrix is only a function of $\tfg_{2}$ and $\lambda^{2}$. One can calculate the eigenvalues (or the critical exponents) from the above matrix as
\newpage
\beq\label{exteigen}
\Lambda_{0}\=-d, \\
\Lambda_{1,2}\=-\frac{1}{2\Delta_{*}}\ll 28 g_{2}^{*}+(6-d)\Delta_{*}\pm \l 688g_{2}^{*2}-40(d-2)^{2}g_{2}^{*}\Delta_{*} +(d-2)^{2}\Delta_{*}^{2}\r^{1/2}\rr.
\eeq
From the spike analysis, it is evident that  $g_{2}^{*}\sim (1+\tlam)$. Using this relationship, it can be readily observed that the critical exponents remain independent of the non-local coupling $\tlam$ or the same as the one in the standard local theory. This holds true even for higher-order truncations and in fractional dimensions. So we are looking at the same fixed point as though the non local coupling changes its values.  

The nonlocal coupling affects the phase transitions (see Fig. \ref{extphase}). The dotted red line in Fig. \ref{extphase} represents the separatrix for local theory ($\tlam =0$), and the point "P" moves to the symmetry phase for a local theory. However, once we turn on the nonlocal parameter, the flow diagram changes (dotted blue and green lines show the separatrix for different values of $\tlam$), and for $\tlam\geq 0.2$, the point $``P"$ moves to the broken phase. So, one can say that the nonlocal parameter induces symmetry breaking.
\vspace{-1em}

\subsection{Beyond the polynomial truncation}\label{secbeyondpoly}

It is interesting to study the evolution of the flow without resorting to polynomial truncations. 
Due to the nonlinearity of the functional flow equation already in the LPA approximation, we shall keep $\lambda$ constant and explore the impact of the $\lambda^2$ term in the renormalized flow in the limit $k \rightarrow 0$. 
The question we would like to address is whether the presence of a nonlocal term affects the convexity properties of the potential as $k \rightarrow 0$. 
To investigate this problem, it is convenient to recast Eq. \ref{unscalukflow} as a flow equation for the threshold function
\begin{equation}
w = \biggl(1 + \frac{\lambda^2}{k^4} + \frac{u_k^{''}(\phi)}{k^2} \biggr)^{-1} \, ,
\end{equation}
so that, after some algebra, one obtains the following quasi-linear parabolic equation in terms of the variable $\phi=x$
\begin{equation}
\label{wd}
\dot{w} = 
-2 w  \left(1-w+\lambda ^2 e^{4 t} w\right)
+e^{-t(d -2) } \left(1-\lambda ^2 e^{4 t}\right) w'' w^2,
\end{equation}
where a dot denotes a derivative with respect to the (RG) time $t$, and $e^{-t} := k$ and we have set $\mu_d=1$.
In this formalism the physical limit $k \rightarrow 0$ is therefore obtained as the large-time behavior of Eq.~(\ref{wd})~\cite{BONANNO200436,CAILLOL2012854,PhysRevE.76.031113}.
Equation~(\ref{wd}) can be interpreted as a modified heat equation with a field-dependent thermal diffusivity coefficient
\newpage
\begin{equation}
e^{-t(d-2)} w^2 \left( 1 - e^{4t} \lambda^2 \right) \, ,
\end{equation}
whose sign is, however, positive definite only for $\lambda^2 \le 0$. 
When this condition is violated, Eq. \ref{wd} does not define a well-posed initial value problem and the solution is not guaranteed to exist. 
As a consequence, the flow can become singular at a finite and positive value of $t$, signaling that it is not always possible to continue the evolution towards the infrared for the nonlocal theory defined by Eq. \ref{mainL} for $\lambda^2>0$ at least in this truncation.  This situation must be contrasted with the local theory, $\lambda = 0$, where the limit $t \rightarrow \infty$ is well defined below the critical line, both in $d = 3$ and $d = 4$.
On the other hand, if we try to analitically continue our equation to  $\lambda^2 < 0$ values, the numerical integration does not lead to a singular flow in the infrared, and in this case the system reaches a trivial, high-temperature fixed point for $r>0$ and an ordered phase for initial conditions below the critical line. 
We have therefore numerically integrated Eq. \ref{wd} using the method of lines, as implemented in \textit{Mathematica}, as already discussed in Ref.~\cite{Bonanno:2022edf} and explicitly checked this scenario considering a bare theory with a potential $r x^2/2 +g x^4$ potential. 
\begin{figure}
 \centering
 \includegraphics[height=10cm, width=13cm]{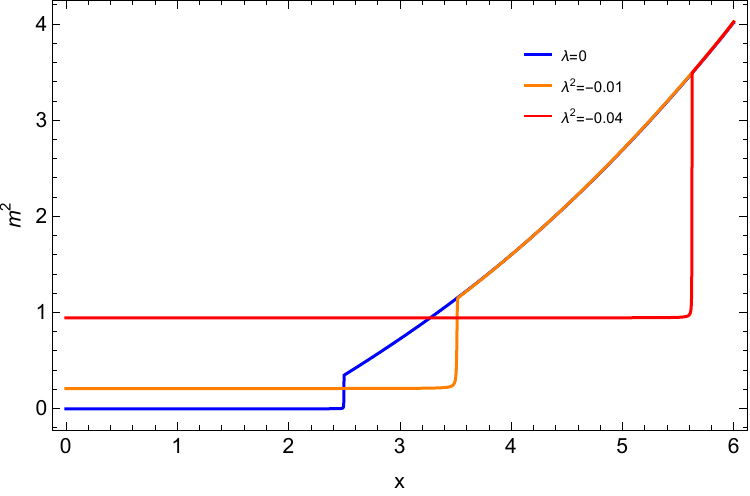}
 \caption{The generalized ``mass'' $m^2 = u''(x, t)$ obtained from the integration of the flow in the limit $t \rightarrow \infty$ for $r = -0.3$ and $g = 0.01$. 
 The solid blue line corresponds to $\lambda^2 = 0$, the orange line to $\lambda^2 = -0.01$, and the red line to $\lambda^2 = -0.04$. }
 \label{flowinte}
\end{figure}
The results are shown in Fig.~\ref{flowinte} for $d = 3$, where the recovery of the convexity properties of the effective blocked potential in the limit $t \rightarrow \infty$ is clearly visible, for $r=-0.3$ and $g=0.01$.
In contrast, it turns out that  it is not possible to extend the integration to $k=0$ ($t\rightarrow \infty$) for $\lambda^2 > 0$ as a singularity always appears at some  $t_{sing}<\infty$.   It would be interesting to study under which set of initial data it is possible to produce non-singular flow in the large $t$ limit. The question is certainly important but because of the mathematical and numerical difficulties in treating this case, we postpone this analysis to a future investigation and focus in this analysis only on the critical properties of the theory. 
 
\subsection{Including the flow of nonlocal coupling}\label{nlflowsec}
Once we consider the non local coupling is also flowing,  in the LPA, along with the flow equation for the potential (Eq. \ref{scaledflowu}) we have $\p_{t}\tlam = -4\tlam$, which at the fixed point gives $\tlam =0$,  the standard local theory. So inorder to capture the effects of flowing nonlocal coupling we move to the next-order approximation by incorporating scale-dependent wavefunction renormalization ($Z_{1}$) as
\beq\label{ansatz}
\gak = \int_{x}\frac{Z_{1}}{2}\l\p\phi\r^{2}+\frac{\lambda^{2}}{2}\phi\frac{1}{-\Box}\phi + \uk\l\rho\r.
\eeq
Including the wavefunction renormalization we rewrite our cotoff regulator function as 
\beq\label{regulator}
\rk = \left[ Z_{1}\l k^{2}-q^{2}\r+\lambda^{2}\l\frac{1}{k^{2}}-\frac{1}{q^{2}}\r\right]\theta\l k^{2}-q^{2}\r.
\eeq
Using our ansatz (Eq. \ref{ansatz}) and the regulator function (Eq.\ref{regulator}) in \weq one can obtain the  renormalization group flow equation for the parameters as (see Appendix \ref{App1})
\beq\label{unscalukflow2}
\p_{t}\lam2\= 0, \quad \hspace{0.5em} 
\p_{t}\z1 = \mu_{d}k^{d+2}\l \uk'''\r^{2}\l\z1 -\frac{\lam2}{k^{4}}\r^{2}\l \z1 k^{2}+\frac{\lam2}{k^{2}}+\uk''\r^{-4},\\
\p_{t}\uk\= -\mu_{d}k^{d+2}\l \z1-\frac{\lam2}{k^{4}}+\frac{\p_{t}\z1}{d+2}\r \l \z1 k^{2}+\frac{\lam2}{k^{2}}+\uk''\r^{-1},
\eeq
where $\mu_{d}^{-1}= \l 4\pi\r^{d}\Gamma\l d/2+1\r$.  In Eq.  \ref{unscalukflow}, we must specify the field configuration used to calculate $\p_{t}\z1$,  which we take at the minimum of the potential ($\rho =\kappa$). We need the dimensionless flow equation for the parameters to analyse the fixed point structure of the theory.  Here we have two natural choices to choose the dimension of the parameters. First is similar to the standard local theory, where we choose $Z_{1}$ as dimensionless and absorbed into the field where the scaling is given as
\beq\label{localscaling}
\phi \= \sqrt{\frac{\mu_{d}}{\z1}}k^{\frac{d-2}{2}}\tphi, \quad u = \mu_{d}k^{d}\tu, \quad z=\tilde{z},\quad \lam2 = k^{4}\tlam.
\eeq
This leads to the definition of $\tilde{J}_{nl}=\tlam/\tilde{\z1}$ along with which we can write the scaled flow equation as
\beq
\p_{t}\tjnl \=  \l 4-\eta\r \tjnl \text{\; with}\quad
\eta \equiv -\frac{\p_{t}\z1}{\z1}= \l \tu'''\r^{2}\l 1-\tjnl\r^{2}\l 1+\tjnl+\tu''\r^{-4},\\
\p_{t}\tu \= d\tu -\frac{1}{2}\l d-2\r\tphi \tu'-\l 1-\tjnl -\frac{\eta}{d+2}\r\l 1+\tjnl+ \tu''\r^{-1}.
\eeq
Looking at the flow equation for $\tjnl$, one can see the possible solutions are $\tjnl=0$ and $\eta=4$.  The $\tjnl=0$ corresponds to the standard local theory with known fixed points in the corresponding dimensions \cite{Codello_2012}.  The $\eta=4$, can't be studied properly using this scaling, for which we need the next scaling (see section \ref{arbsigma} for more details).

In order to capture the effect of non local coupling we consider the second scaling where we treat $\lambda^{2}$ as dimensionless and absorb it into the field redefinition (\cite{PhysRevE.92.052113, Defenu_2020}) which gives the dimensionless parameters as 
\beq\label{scale}
\phi = \l\frac{\mu_{d}}{\tlam}\r^{\frac{1}{2}}k^{\frac{d+2}{2}}\tilde{\phi},\quad
\uk = \mu_{d} k^{d}\tu,\quad
\z1 = \frac{\tz}{k^{4}},\quad
\lam2 =\tlam.
\eeq   
This also leads to the definition of non local parameter $\tilde{J}_{l} =\tz/\tlam$.  Along with $\p_{t}\tlam = 0,$ one can write the flow equation for the scaled parameters as
\beq\label{nlscflow}
\p_{t}\tj \= -4\tj +\l \tu'''\r^{2}\l \tj-1\r^{2}\l 1+ \tj+\tu''\r^{-4},\\
\p_{t}\tu \= d\tu -\l\frac{d+2}{2}\r\tphi\tu' -\l \tj-1+\frac{\p_{t}\tj -4\tj}{d+2}\r\l 1+\tj +\tu''\r^{-1}.
\eeq
In the limit $\tj\rightarrow 0$, the fixed point equation can be map to that of the local theory with an effective negative spacetime dimension, as expected from the effective dimension approach \cite{Defenu_2020,PhysRevE.92.052113}. The limit 
$\tj\rightarrow 0$ corresponds to a regime where only nonlocal interactions remain, as one would expect in the far infrared limit. One can solve the above equations (Eq. \ref{nlscflow}) simultaneously as follows,  first solve $\p_{t}\tu =0$ with $\tj =0$, then use the solution of $\tu$ in the flow equation for $\tj$. Repeat this until we obtain a convergent solution for $\tj$.  We can proceed in two ways, one is the standard spike analysis (or shooting from the origin) \cite{Codello_2012,Bridle2014}, where we parametrize the initial conditions using $\tu^{(1)}(\tphi=0)=0$ and $\tu^{(2)}(\phi=0)= s$ and we tune $``s"$ such that we obtain a global solutions to the differential equation.  In the second method we consider the asymptotic form of $\tu$ for large values of $\tphi$, which is a functional of arbitrary parameter $``A"$ and we tune $``A"$ to obtain the global solutions which satisfy $\tu'(\phi=0)=0$ \cite{Bridle2014} (See Appendix \ref{appshoot}).

\begin{figure}[h!]
\centering
\includegraphics[height=9cm, width=13cm]{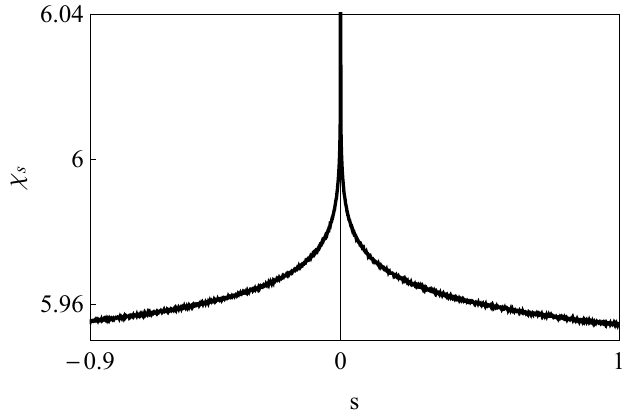}
\caption{Spike analysis plot for Eq. \ref{uflowwithlog}. Here $s=u''\l\phi=0\r$ and $\chi_{s}$ represents the value of $\chi$ where the solution diverges. For the plot we have further scaled the field value as $\tilde{\chi}=10^{-2}\chi$. The plot remains the same for $d=2.2,2.4,2.6,2.8,3,4$. }
\label{spike}
\end{figure}

Here we do the spike analysis starting from $J=0$. In spike analysis we need the values of the field $\phi$ where the solution diverges for a choice of $s$.  As the solution goes slowly to infinity (see Eq. \ref{shootingasym}) the singularity occurs for very large values of $\phi$ which make it hard to obtain the spike. One can circumvent this issue by rescaling the field as $\chi = \log\l 1+\tphi\r$. With this rescaling one re write the flow equation for the scaled potential as
\beq\label{uflowwithlog}
d u -\frac{1}{2}\l d+2\r \l 1- e^{-\chi}\r u'+\ll 1+e^{-2\chi}\l u''-u'\r\rr^{-1}\= 0,
\eeq
where now $u'$ is the derivative w.r.t. $\chi$. With Eq. \ref{uflowwithlog}, we can do the standard spike analysis and the result is shown in Fig. \ref{spike}, from which we can see that there only exist the Gaussian fixed point corresponds to $s =0$.  As the Gaussian fixed point corresponds to constant potential, the flow equation for $\tj$ gives $\tj=0$.  We have repeated the analysis for different values of $d$ including the fractional dimensions where the behavior remains the same.
We have further confirmed the results using the polynomial truncation of the potential with the following ansatz:
\beq\label{pot}
\tu \= \sum_{n=2}^{N}\frac{\tg_{n}}{n!}\l \trh-\tk\r^{n}.
\eeq
Note that we have changed the variable from $\tphi$ to $\trh=1/2\tphi^{2}$. As a first order in approximation one can take the lowest truncation i.e., $N=2$ in Eq. \ref{pot}, then the flow equations are
\beq\label{N2}
\p_{t}\tlam\= 0,\\
\p_{t}\tj \= -4\tj +18\tk\tg_{2}^{2}\l \tj-1\r^{2}\l 1+ \tj+2\tk\tg_{2}\r^{-4},\\
\p_{t}\tg_{2} \= -(d+4)\tg_{2} -18\tg_{2}^{2}\l \tj-1+\frac{\p_{t}\tj -4\tj}{d+2}\r\l 1+\tj +2\tk\tg_{2}\r^{-3},\\
\p_{t}\tk \= \l d+2\r\tk-3\l \tj -1+\frac{\p_{t}\tj -4\tj}{d+2}\r\l 1+\tj +2\tk\tg_{2}\r^{-2},
\eeq
where $\tilde{\kappa}=\tilde{\rho}_{0}$ and we obtain the flow equation for $\tilde{\kappa}$ by scaling Eq. \ref{tkappa}. We find the fixed points of the theory by solving Eq. \ref{N2} for different dimensions ($2< d\leq 4$), including fractional ones. There is only a Gaussian fixed point, corresponding to $\tj^{*} = 0, \tg_{2}^{*}=0$ and $\tk^{*} = -3/(d+2)$. Even after increasing the truncation in Eq. \ref{pot} up to $N=13$, only the Gaussian fixed point remains. There exist spurious fixed point which eventually converges to Gaussian fixed point or to the infinite solution. 

We can analyze the stability of this non local Gaussian fixed point by calculating the eigenvalues of the stability matrix.However in the stability matrix,  at higher truncation ($N>2$), the flow equation for $\kappa$ causes divergences for Gaussian fixed point. So we take $\kappa=0$ with $n$ starts from zero in Eq. \ref{pot}. In  this ansatz the non local Gaussian fixed point corresponds to $\tj=0, \tg_{0}=-1/d$ all other $\tg_{n}=0$, which agrees with the spike analysis.  With this truncation we obtain that all the eigenvalues are positive except for one along the $\tg_{0}$ directions with eigenvalue $\theta=-d$. We have also calculated the eigenvalues of the standard local fixed point, which always has more number of negative eigenvalues (see section \ref{arbsigma}). Also, as we are looking a Gaussian fixed point we can check the stability of the fixed point in a more general way without assuming any form for the potential \cite{PhysRevD.62.027701,HASENFRATZ1986687}. One can linearise the potential around the Gaussian fixed point as $\tu = \tu^{*}+ e^{\theta t}v$, where $\tu^{*}$ is the constant solution at the Gaussian fixed point and $v$ is the small change, which is a function of $\phi$ alone. Then one can find the linearised flow equation for $v$ from Eq. \ref{nlscflow} which gives general solution as
\beq
v\l \phi\r \= e^{-\phi^{2}}\ll C_{1}\; H_{-\nu}\l \phi\r + C_{2}\;\; {}_1F_{1}\l \frac{\nu}{2},\frac{1}{2},\phi^{2}\r \rr,
\eeq
where $\nu= \l 2+3d+2\theta\r/\l d+2\r$. If we further restrict to the polynomial form of the potential, which fixes $C_{1}=0$ and $\nu$ to a positive odd integer which gives $\theta >0$ as obtained from the polynomial truncation. So one can conclude that the non local Gaussian fixed point is the IR stable fixed point with this order of approximation.
\newpage
\section{Local theory}\label{seclocal}
As we mentioned before, the non local theory we considered above can be obtained from a local theory with the action
\beq
S \= \int_{x}\frac{1}{2}\l\p\phi\r^{2}+\frac{1}{2}\l\p\psi\r^{2}-\lambda\phi\psi+u\l\phi\r.
\eeq
We can obtain the non-local theory by path integrating over the $\psi$ field and taking $\lam2\rightarrow -\lam2$ (see \cite{Narain_2019,PhysRevD.99.125012} for discussions on tachyons and complex poles in theory). Therefore, we expect the fixed point structures of the non-local theory to be re-derived from the local theory, which we carry out in this section. To stay consistent with the approximations made earlier, we use the following ansatz for the effective action:
\beq\label{gamlocal}
\gak = \int_{x}\frac{\z1}{2}\l\p\phi\r^{2}+\frac{\zz}{2}\l\p\psi\r^{2}-\lambda\phi\psi + u\l\phi\r.
\eeq 
We need to select a regulator function to obtain a closed-form expression for the flow equations. In this model, we use the standard Litim cutoff \cite{LITIM200092} with
\beq\label{lolitim}
\rkk = \begin{pmatrix}
R_{1} & 0\\
0& R_{2}
\end{pmatrix},
\eeq
where $R_{i} = Z_{i}\l k^{2} -q^{2}\r\theta\l k^{2} - q^{2}\r$.  Note that as we have two fields, we also need to take trace over field indexes in the Wetterich equation (Eq. \ref{weq}). The associated two point vertex function is given as 
\beq
\gg2\=\begin{pmatrix}
\ga2_{11} & -\lambda\\
-\lambda & \ga2_{22}
\end{pmatrix}, \quad \text{with }\hspace{0.3cm} \ga2_{ii} = Z_{i}q^{2} + \mu_{i},
\eeq
where $\mu_{1}=\uk^{(2)}$ and $\mu_{2}=0$.  In this model we can re write \weq as
\beq\label{loweq}
\p_{t}\ggk \= \frac{1}{2}\tr\l \p_{t}\rkk\l \gg2+\rkk\r^{-1}\r.
\eeq
Note that the two-point vertex function does not depend on $\psi$, making the right-hand side of Eq. \ref{loweq} a function of $\phi$ only. First, let us focus on the flow equation for the two-point vertex function, which will provide us with the flow equations for the non-local coupling and the wavefunction renormalization. The flow equation for the two-point vertex function is 
\newpage
\beq\label{lo2p}
\p_{t}\gg2\=\begin{pmatrix}
p^{2}\p_{t}\z1 +\p_{t}\mu_{1} & -\p_{t}\lambda\\
-\p_{t}\lambda & p^{2}\p_{t}\zz
\end{pmatrix}.
\eeq
To derive the flow equations for $\lambda$  and $\z1$, we need to take the second derivative of Eq. \ref{loweq} with respect to the fields. However, since the right-hand side of Eq. \ref{loweq} is independent of $\psi$, all derivatives with respect to $\psi$ vanish. Thus, we conclude that $\p_{t}\lambda = 0$ and $\p_{t}Z_{2} = 0$ are consistent with the results from the previous section. To find $\p_{t}\z1 $, we focus on $\p_{t}\Gamma^{(2)}_{11}$, which is the second derivative of Eq. \ref{loweq} with respect to $\phi$. Proceeding in the same way as for the non-local theory, we can show that $\p_{t}\z1$ matches the expression for the non-local theory in Eq. \ref{unscalukflow} (see Appendix \ref{App12}). Next, we calculate the flow equation for the other parameters from Eq. \ref{loweq}, we can proceed in two ways: by performing the momentum integral and then taking the trace of the matrix, or by taking the trace first and then performing the momentum integral. Both methods yield the same result as
\beq
\p_{t}\uk\=-\mu_{d}k^{d+2}\l 2\z1 k^{2}+\uk^{(2)}+\frac{k^{2}\p_{t}\z1}{d+2}\r\l k^{2}\l\z1 k^{2}+\uk^{(2)}\r-\lam2\r^{-1}.
\eeq
where we took $Z_{2}=1$. To compare the result with the non-local one we take $\lam2\rightarrow -\lam2$, which gives
\beq
\p_{t}\uk\=-\mu_{d}k^{d} -\mu_{d}k^{d} \l \z1 k^{2}-\frac{\lam2}{k^{2}}+\frac{k^{2}\p_{t}\z1}{d+2}\r\l \z1 k^{2}+\uk^{(2)} +\frac{\lam2}{k^{2}}\r^{-1} .
\eeq
This aligns with the flow equation for the non-local $\uk$ (see Eq. \ref{unscalukflow}) up to a constant. This constant is related to the normalization of the Jacobian when performing the path integral over $\psi$ to derive the non-local action.
\vspace{-1em}
\section{Interactions of the form $\bigl(-\Box \bigr)^{\sigma/2}$}\label{arbsigma}
\begin{figure}
\centering
\includegraphics[height=9cm, width=13cm]{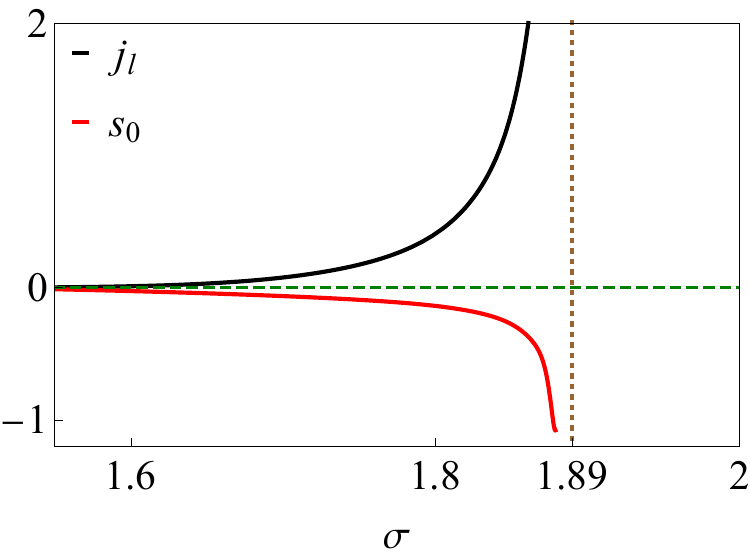} 
\caption{Plotted the values of $\tj$, $\kappa$ and $\tg_{2}$ as function of $\sigma$ for $d=3$. The purple color show the non local fixed point and the green dotted line is for the nonlocal Gaussian fixed point.}
\label{sigmaspike11}
\end{figure}

\begin{figure}
\centering
\includegraphics[height=9cm, width=13cm]{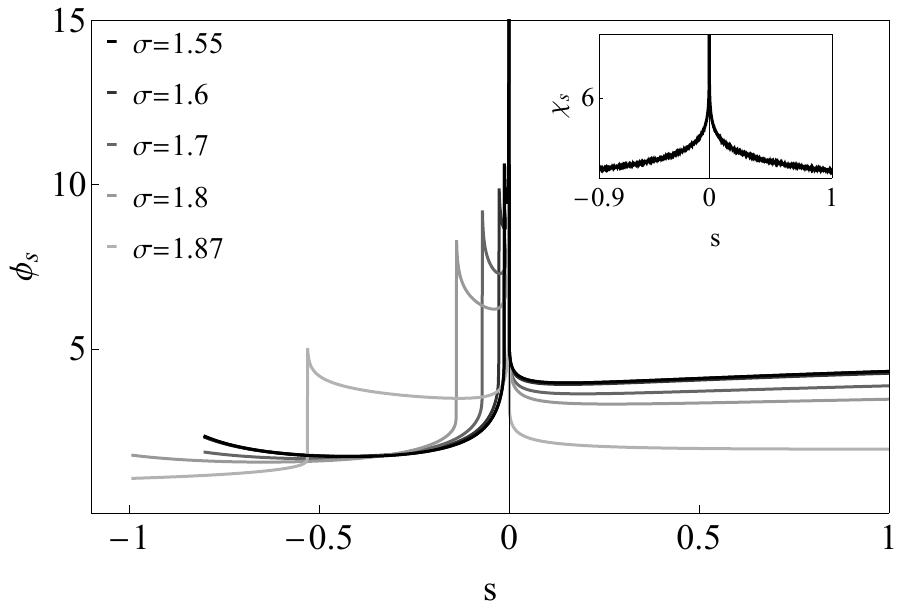}
 \caption{The spike analysis of Eq. \ref{sigmaflow} for different values of $\sigma$ for $d=3$.  The single non Gaussian fixed point corresponds to a non local fixed point which merges with the NL GFP at $\sigma=1.5$. Inset shows the spike plot for $\sigma=-2$, where note the change in y-axis as we scale the field. We gets qualitatively the same plot for all $-2<\sigma<1.5$.}
\label{sigmaspike1}
\end{figure}

In the previous section, we observed that the study of renormalization group structure of a non local theory is consistent with the equivalent local theory.  We now extend this analysis to a broader class of non-locality by considering a theory with the action
\beq\label{sigmaS}
S = \int_{x}\frac{\z1}{2}\l\p\phi\r^{2}+\frac{\lambda^{2}}{2}\phi \l -\Box\r^{\frac{\sigma}{2}}\phi + U\l\rho\r.
\eeq
The fractional power of $\Box$ can be understood in momentum space as $p^{\sigma}$. For positive values of $\sigma$, this model corresponds to the study of weak long-range interactions, as discussed in \cite{Defenu_2020, PhysRevE.92.052113}. One can repeat the analysis in section \ref{secbeyondpoly} for $Z_{1}=1$ and see that the flow equation is not singular for $\sigma>0$. Here, we extend the critical property analysis to negative values of $\sigma$, focusing specifically on the consistency with $\sigma = -2$ , which was examined in the previous section (section \ref{model}). One can proceed in the same as in the previous section using the similar ansatz for effective average action and the regulator by replacing $-\Box$ with $\l -\Box\r^{\sigma/2}$, where now the regulator takes the form
\beq\label{regulatorsigma}
\rk = \left[ Z_{1}\l k^{2}-q^{2}\r+\lambda^{2}\l k^{\sigma}-q^{\sigma}\r\right]\theta\l k^{2}-q^{2}\r.
\eeq
Similar to the previous section we have two natural choices for the dimensionless couplings. As we are more interested in the negative values $\sigma$, we first uses the one with constant nonlocal coupling.  As the flow equation is analytic in $p$, we can obtain $\p_{t}\tlam=0 $ and along with the scaling Eq. \ref{scale} one can write the flow equations as
\beq\label{sigmaflow}
\p_{t}\tj \= \l \sigma-2\r\tj +\frac{1}{4}\l \sigma+2\tj\r^{2}\l \tu'''\r^{2}\l 1+ \tj+\tu''\r^{-4},\\
\p_{t}\tu \= d\tu -\frac{1}{2}\l d-\sigma\r\tphi\tu' -\l \tj+\frac{\sigma}{2}+\frac{\p_{t}\tj +(\sigma-2)\tj}{d+2}\r\l 1+\tj +\tu''\r^{-1}.
\eeq
Similar to the above section now one can solve Eq. \ref{sigmaflow} for specific values of $\sigma$ and $d$ using the spike analysis as shown in the Fig. \ref{sigmaspike1}.  First solve the flow equation for $\tu$ using spike analysis with $\tj =0$ and use the solution in the flow equation for $\tj$ and iteratively solve the system until we obtain a convergent value for $\tj$. In Fig. \ref{sigmaspike11}, we have plotted the convergent value for $\tj$, and $s_{0}=\tu''(\tphi=0)$ as a function of $\sigma$.  We can see that $\tj$ and $s_{0}$ diverges as we approaches $\sigma^{*}<2$, where the $\tlam$ vanishes and theory is determined by the standard local theory. Also, as we move towards $\sigma=d/2$, the fixed point merges with the non local Gaussian fixed point.  This can also be seen the spike plot in Fig. \ref{sigmaspike1}. Similar conclusions are obtained in \cite{Defenu_2020,PhysRevE.92.052113} using polynomial truncation from which we know that $\sigma^{*}=2-\eta$ as obtained in \cite{PhysRevB.8.281} using perturbative analysis. We also verified the above statement using spike analysis, where we calculated $\eta=0.109$ for the Wilson Fischer fixed point which make $\sigma^{*}\sim 1.89$ as one can see in Fig.\ref{sigmaspike11}
In the standard local theory, the asymptotic behaviour for the large values of $\tphi$ is dominated by the scaling part of the equation (first terms in the flow equation for $\tu$ in Eq. \ref{sigmaflow}).  For an arbitrary values of $\sigma$ and $d$, one can solve the scaling part to get $\tu \sim \tphi^{(2d)/(d-\sigma)}$ which dominates in the $\tphi\rightarrow\infty$ limits for $0<\sigma<d$. Or for large values of $\tphi$, one can neglect the contribution from the quantum corrections, the non linear contribution (the last term in the flow equation for $\tu$ in Eq. \ref{sigmaflow}), as $\tu''>>1$. However, all this behaviour changes for $\sigma<0$, the non linear part also contribute to the asymptotic behaviour and the solutions goes slowly to infinity. That shows, in the strong long range regime ($\sigma<0$) \cite{RevModPhys.95.035002}, the behaviour in $\tphi\rightarrow\infty$ regime is also contributed by $\tphi\sim 0$ fields.  Because of this reason one can't directly do the spike analysis for $\sigma<0$, as one has to scale the field appropriately inorder to include $\tphi\rightarrow\infty$ regions as discussed around Eq. \ref{uflowwithlog}. 

The flow equation (Eq. \ref{sigmaflow}) reduces to standard local theory  for $\sigma=2$ in the $\tj\rightarrow 0$ limit (at $\sigma=2$ both $\tlam$ and $\z1$ act as wave function renormalization and $\tj\rightarrow 0$ means $\z1\rightarrow 0$ and $\tlam$ act as wavefunction renormalization) and we recover the spike plots for the local theory (also see the effective dimension method as discussed in \cite{PhysRevE.92.052113}). As we decrease the $\sigma$ value from $\sigma =2$ to $\sigma=d/2$, the nonlocal non-Gaussian fixed moves towards the nonlocal Gaussian one and merges with it at $\sigma =d/2$.  After $\sigma=d/2$, there only exist the Gaussian fixed point and inorder to see this for $\sigma<0$, we need to do the spike analysis after rescaling the field value (see the plot in the inset of Fig.\ref{sigmaspike1}).  The behaviour remains in the same for arbitrary dimensions $2\leq d\leq 4$ including the fractional dimensions.

\begin{figure}
     \centering
      \begin{subfigure}{0.5\textwidth}
         \centering
         \includegraphics[scale=0.51]{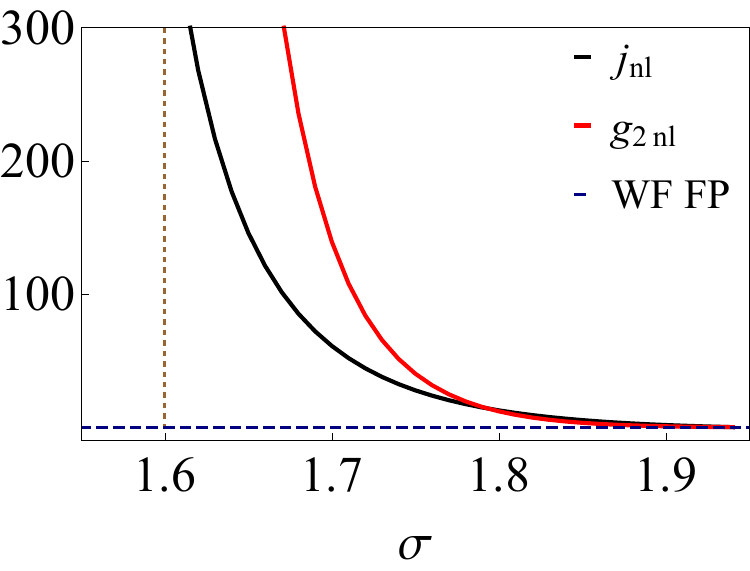} 
     \end{subfigure}
     \begin{subfigure}{0.49\textwidth}
         \centering
      \includegraphics[scale=0.42]{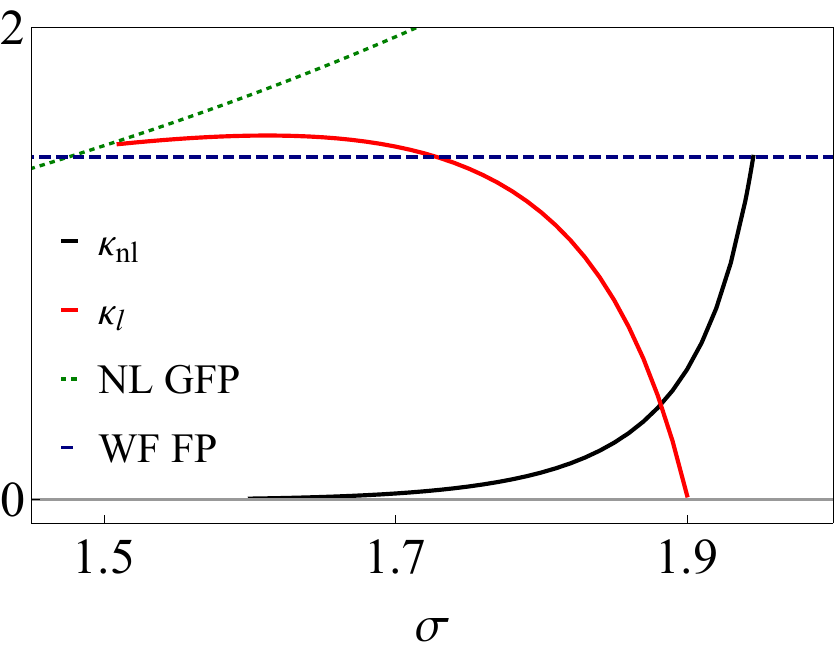}
     \end{subfigure}
      \caption{Plotted the values of $j_{nl}, g_{2nl}$ and $\kappa$ as function of $\sigma$ for $d=3$. The $j_{nl}$ and $g_{2nl}$ diverges near $\sigma\sim d/2$ as the fixed point merges with the nonlocal Gaussian fixed point.}
 \label{sigmatrunc1}
 \end{figure}

\begin{figure}
\centering
 \includegraphics[height=9cm, width=15cm]{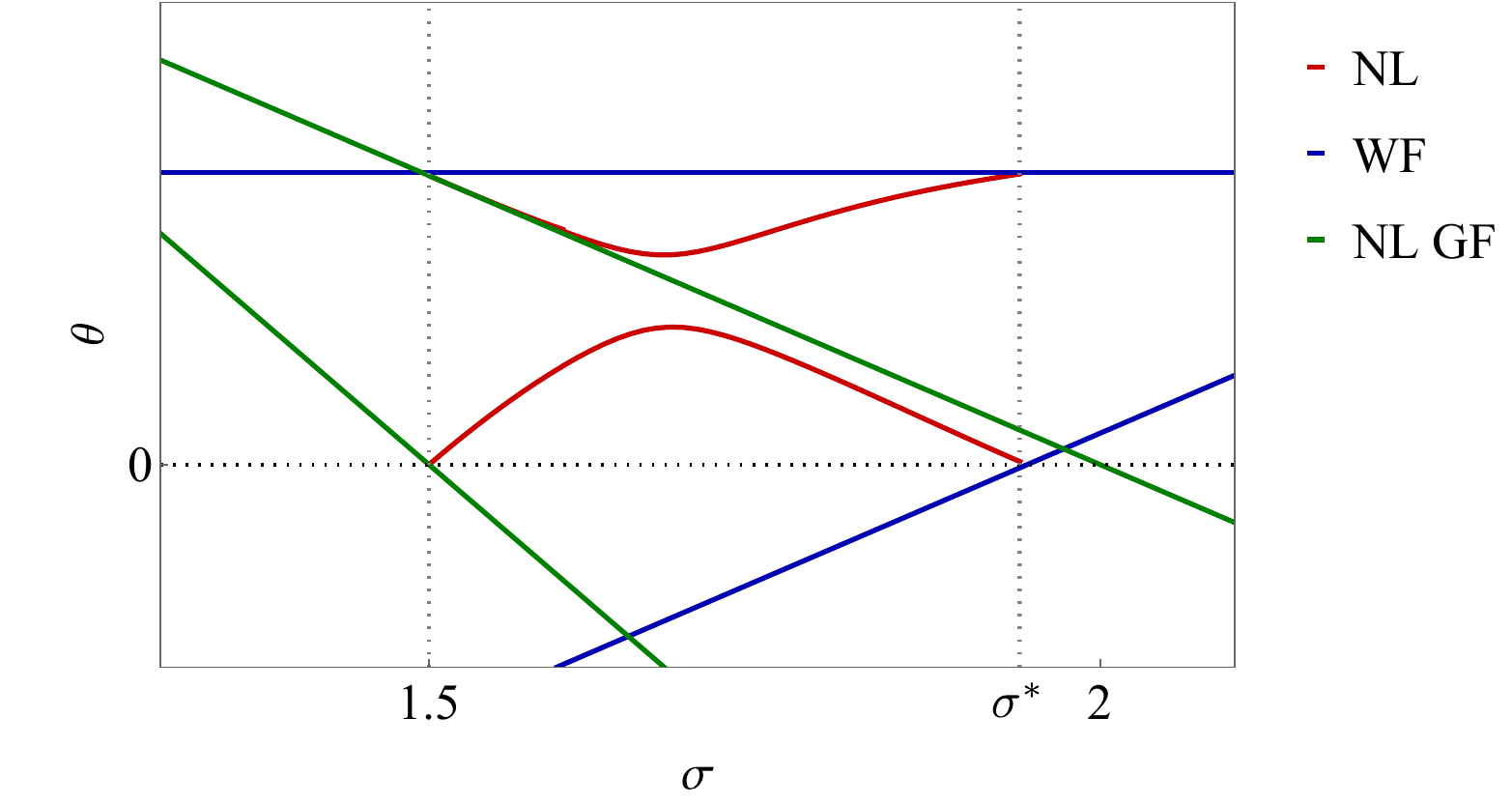} 
 \caption{The plot shows two eigenvalues ($\theta$) of the RG stability matrix as a function of $\sigma$ in three dimension for the local WF fixed point (blue lines), the NLG fixed point (green lines) and the NL fixed point (red lines). WF is stable in $\sigma> \sigma*$ with two positive eigenvalues, NL is stable in $1.5<\sigma<\sigma*$ as one of the positive eigenvalues of WF turns negative for $\sigma<\sigma*$ and NL GF is stable for $\sigma<1.5$ with two positive eigenvalues. As we move from left to right, the IR stable theory becomes more local.}
  \label{allexponenets}
\end{figure}

Similar to section \ref{nlflowsec}, one can also solve Eq. \ref{sigmaflow}, using the polynomial truncation as in Eq. \ref{pot}, which for $N=2$ gives
\beq\label{sigmatrun}
\p_{t}\tj \=\l\sigma-2\r\tj +\frac{9}{2}\tg_{2}^{2}\kappa \l \sigma+2\tj\r\l 1+\tj +2\kappa\tg_{2}\r^{-4}, \\ 
\p_{t}\tg2 \=\l 2\sigma-d\r\tg_{2}-18\tg_{2}^{2}\l\tj+\frac{\l \sigma-2\r\tj}{d+2}+\frac{\sigma}{2}\r\l 1+\tj +2\kappa\tg_{2}\r^{-3}, \\
\p_{t}\tk \= \l d-\sigma\r\kappa -3\l\tj+\frac{\sigma}{2}+\frac{\l \sigma-2\r\tj}{d+2}\r \l 1+\tj +2\kappa\tg_{2}\r^{-2}.
\eeq
One can solve Eq. \ref{sigmatrun} numerically for different values of $\sigma$ and $d$ to obtain qualitatively similar plot as in Fig. \ref{sigmaspike11} as obtained in \cite{PhysRevE.92.052113,PhysRevE.92.052113}. 
Now we consider the other scaling where we take $\z1$ as dimension less (see Eq. \ref{localscaling}), which is the best to study the transition at $\sigma^{*}$. With this scaling one can write the scaled flow equation as
\beq\label{sigmalocal}
\p_{t}\tjnl \=  \l 2-\sigma -\eta\r \tjnl \text{\; with}\quad
\eta \equiv -\frac{\p_{t}\z1}{\z1}= \frac{1}{4}\l \tu'''\r^{2}\l 2+\sigma\tjnl\r^{2}\l 1+\tjnl+\tu''\r^{-4},\\
\p_{t}\tu \= d\tu -\frac{1}{2}\l d-2+\eta\r\tphi \tu'-\l 1+\frac{\sigma}{2}\tjnl -\frac{\eta}{d+2}\r\l 1+\tjnl+ \tu''\r^{-1}.
\eeq
We can do similar analysis using truncation like Eq. \ref{pot} and values of $\tjnl, \kappa$ and $\tg_{2}$ as a function of $\sigma$ is given in Fig. \ref{sigmatrunc1}.
From the plot Fig. \ref{sigmatrunc1} and Fig. \ref{allexponenets}, on can see that as we moves towards $\sigma^{*}$, the non local fixed point emerges from the standard Wilson Fischer fixed point. The plot diverges near $\sigma=1.5$ ($d/2$) as the non local fixed point merges with the non local Gaussian fixed point, which we studied using the first scaling. The standard local theory corresponds to $\tjnl=0$ in the flow equation for $\tjnl$ in Eq.  \ref{sigmalocal} and $\eta=2-\sigma$ solution corresponds to the non local fixed point.  In our analysis of $\sigma=-2$, we mention that we are not taking the $\eta=4$ case, which we can understand as invalid. 

Now one can study the stability of different fixed points in different regions of $\sigma$ using the eigenvalues of the stability matrix. The critical exponents of different fixed points at different values of the $\sigma$ for $N=2$ truncation for $d=3$ is plotted in Fig. \ref{allexponenets}. One can see that for $\sigma> \sigma^{*}$ the Wilson Fisher fixed point is the IR stable one with less number of negative eigenvalue. As we cross the $\sigma^{*}$,  one of the positive eigenvalue of the WF turns negative, which make the new nonlocal fixed point as stable one.  As we cross $\sigma<d/2$, then the non trivial non local fixed point merges with the NL GF and it is the stable one in the region.  Note that NL GF corresponds to theory $\z1\rightarrow 0$  and non zero value of $\tlam$, however the standard local GF corresponds to $\tlam\rightarrow 0$ with non zero $\z1$. As expected these fixed points merges at $\sigma=2$.

Note that as discussed before inorder to obtain the critical exponents for both local and non local Gaussian FP with $N>2$, we use the polynomial truncation with zero vacuum expectation value  as
\beq
\tu\l \trh\r \= \sum_{n=1}^{N}\frac{g_{n}}{n!}\trh^{n}.
\eeq
\begin{table}[ht]
\centering
\small
\setlength{\tabcolsep}{20pt} 
\begin{tabular}{|c|c|c|c|c|c|}
\hline
\multirow{2}{*}{$\sigma$ values}
    & \multicolumn{4}{|c|}{No. of negative eigenvalues}
    & \multirow{2}{*}{IR Stable FP} \\
\cline{2-5}
 & GFP & WF & NL & NL GFP & \\
\hline
$\sigma>\sigma^*$             & 2 & 1 & - & $3^{\dagger}$ & WF \\
$d/2<\sigma<\sigma^*$ & 3 & 2 & 1 & 2             & NL \\
$\sigma<d/2$          & 3 & 2 & - & $1^{\dagger}$ & NL GFP \\
\hline
\end{tabular}
\captionsetup{justification=raggedright,singlelinecheck=false}
\caption{Shows the No. of negative eigenvalues for different FP for different ranges of $\sigma$. More negative eigenvalues means FP is unstable in the IR region.  $^{\dagger}$ means as we increase the truncation number changes without changing the IR stable FP (we have more number of negative eigenvalues for $\sigma>\sigma*$ and all eigen values turns positive for $\sigma<0$).}
\end{table}
\newpage
\subsection{Isotropic Lifshiftz fixed point}
In this section, we extend our analysis to $\sigma>2$, resulting in a higher-derivative theory. We compare our findings with those in \cite{BONANNO2015501} on isotropic Lifshiftz fixed point, which is equivalent to $\sigma=4$ in our model. A full analysis of the transition from $\sigma=2$ to $\sigma=4$ may require higher-order approximations, incorporating additional higher-derivative terms. However, a leading-order approximation can be performed by setting $Z=0$ and adopting a scaling where $\lam2$ remains dimensionless and is absorbed through a field redefinition. This approach is equivalent to the local potential approximation (LPA) discussed in \cite{PhysRevE.92.052113}, and the analysis can be carried out using an effective dimensional method \cite{PhysRevE.92.052113}. At this order, one can write the flow equation for the potential as
\beq\label{hdt}
\p_{t}\tu \= \deff\tu -\frac{1}{2} \l \deff- 2\r \tphi \tu - \l1+\uk''\r^{-1},
\eeq
 where $\deff = 2d/\sigma$. The above flow equation (Eq. \ref{hdt}) can be mapped onto a standard local theory in an effective spacetime dimension $\deff$, which allows us to extract key results using known properties of local models. In such local theories, it is well established that a discrete set of fixed points exists for $2<\deff<4$, depending on the precise value of $\deff$. Translating this to the isotropic Lifshitz case, one finds a corresponding discrete set of fixed points for $4<d<8$, identifying $d=4$ and $d=8$ as the lower and upper critical dimensions, respectively, as discussed in \cite{BONANNO2015501}. Based on this correspondence, we conclude that within this approximation, a unique non-Gaussian fixed point exists in the range $6\leq d<8$. So, the use of effective dimensional analysis indeed gives the leading order behaviour of the isotropic Lifshitz fixed point.
\section{Conclusion}\label{seccon}
We studied the fixed-point structure of a single scalar field theory with interactions of the form $\phi \Box^{\sigma/2}\phi$, focusing primarily on $\sigma=-2$ using functional renormalization group (FRG) approach. The theory is a useful toy model for various nonlocal extensions of gravity and provides an effective description of systems with strong long-range interactions. A comprehensive analysis of its phase space structure using the exact renormalization group method is a novel contribution to literature. Our analysis shows that, in the LPA, the flow equation becomes singular for $\lambda^{2}>0$ when $\sigma < 0 $, including the case $\sigma = -2$. We then examined the critical properties and fixed-point structure of the theory. Within the LPA${}^\prime$ approximation, we determine the fixed points and analyse their stability for all $-2 \leq \sigma\leq 2$. We further perform a qualitative examination of the fixed points for $\sigma >0 $ in the LPA.

We obtained closed-form expressions for the flow equations of all parameters, including the wavefunction renormalization within the LPA\textquotesingle, valid in arbitrary spacetime dimensions. Notably, we found that the nonlocal coupling remains scale-invariant, $\p_{t}\lam2 =0$ (see Eq. \ref{unscalukflow} and Eq. \ref{unscalukflow2}). Initially,In LPA, we treated the nonlocal coupling as an external parameter and solved the scaled flow equation using spike analysis (see Fig. \ref{extspike}). At this stage, we found that the fixed-point structure remains identical to that of the local theory, with the nonlocal parameters merely shifting the fixed-point locations. Additionally, we computed the critical exponents—physically measurable quantities— using the truncation method and confirmed that they remain independent of the nonlocal parameter (see Eq. \ref{exteigen}). We also look at the RG flow diagram for the parameters and show that nonlocal coupling affects the phase transition as the nonlocality induces symmetry breaking (Fig. \ref{extphase}). We analyse how the nonlocal term influences the convexity of the potential as $k\rightarrow 0$ within the LPA beyond polynomial truncations. The flow equation is found to become singular for $\lambda^{2}>0$ before the far-infrared limit ($k\rightarrow 0$). Obtaining a smooth flow may therefore require going beyond these approximations, which is computationally demanding. We further investigate the fixed-point structure of the theory.

Next, we considered the flow of nonlocal parameter, which yielded trivial results at the LPA level. This motivated us to move to the next-order approximation by incorporating wavefunction renormalization. We adopted a natural scaling where the nonlocal parameter $\lam2$ remains dimensionless. Under this scaling, we demonstrated—using spike analysis—that the only fixed point present is the nonlocal Gaussian fixed point (NL GFP) (see  Fig. \ref{spike} and appendix \ref{appshoot}, where we independently confirm this result using another method). Additionally, we validated these findings through the truncation method. In addition, we compute the critical exponents using both polynomial truncations and without assuming the polynomial form. These results confirm that the nonlocal Gaussian fixed point is the stable infrared fixed point of the model under consideration (See Fig. \ref{spike}). So, in the low energy limit, only the nonlocal term contributes. 

Moreover, we confirmed our results by rederiving the flow equations for the parameters using the corresponding local theory formulated with two scalar fields. We found that the flow equations match those of the nonlocal theory up to a constant factor. This discrepancy arises from the normalization of the Jacobian in the path integral of the auxiliary field.

Additionally, we explored how infrared physics changes as a function of $\sigma$. We considered two scaling schemes: one in which the wavefunction renormalization remains dimensionless, as in standard local theories, and another in which the nonlocal coupling is dimensionless. We did analysis using the appropriate scaling for different values of $\sigma$. Using both scaling, we solved the scaled flow equation via spike analysis. Starting from $\sigma=-2$, we found that the fixed-point structure remains unchanged up to $\sigma=d/2$, with the NL GFP continuing as the infrared-stable fixed point. At $\sigma=d/2$, a new nonlocal fixed point emerges from the NL GFP, which becomes the IR stable fixed point (see Fig. \ref{sigmaspike1} and Fig. \ref{allexponenets}). This fixed point persists until $\sigma=\sigma^{*}=2-\eta $ ($\eta$ is the anomalous dimension), where it merges with the local fixed point-specifically, in three dimensions, with the Wilson-Fisher fixed point. For $\sigma^{*}\leq \sigma\leq 2$, the local fixed points become the infrared stable one. This behaviour aligns with Sak’s prediction and has been previously demonstrated using the truncation method in \cite{PhysRevE.92.052113}. Additionally, we numerically computed $\sigma^{*}$ using spike analysis. We also repeated the analysis using the truncation method to verify the results. 

A complete understanding of the nonlocal fixed point depends on the choice of scaling. If the wavefunction renormalization is treated as dimensionless, the quantity $J_{nl}$ diverges near $\sigma=d/2$ (see Fig. \ref{sigmatrunc1}), where the nonlocal fixed point merges with the nonlocal Gaussian fixed point. Conversely, if $\lam2$ is taken to be dimensionless, the divergence of $Z$ for $\sigma\geq\sigma^{*}$ cannot be absorbed through a field redefinition, resulting in a divergence of $J_{l}$ near $\sigma=\sigma^{*}$ (see Fig. \ref{sigmaspike11}). So, we use both scaling to understand the nature of nonlocal fixed points. Finally, using the effective dimensional method, we studied the fixed point structure of higher derivative theories, particularly for $\sigma=4$. The model corresponds to the isotropic Lifshitz critical behaviour, and we find that our results align with those in \cite{BONANNO2015501} at this level of approximation.

Along with the results we obtained, we can conclude that the FRG is a suitable tool to study the fixed point structure of a nonlocal theory. The method gives the advantage of keeping the spacetime dimension and the type of nonlocality (value of $\sigma$) as arbitrary. We will further do a careful analysis of the effect of nonlocality on the phase transition, including higher-order approximation, especially studying the phase transition, including the wavefunction renormalization. One can also include the effect of temperature in the same way as in \cite{MARIAN2024139051,PhysRevLett.134.041602}, where we can study the interplay between temperature and nonlocality in phase transition. Another interesting direction is the effect of nonlocality on the Mermin-Wagner theorem by including $N$ scalar fields with $O(N)$ symmetry.
\acknowledgments{It is a pleasure to thank Dario Zappal\'a for useful discussions. We are thankful to CHEP for hosting Alfio Bonanno for a short visit which started a fruitful collaboration. GN would like to acknowledge the startup support from IISc which allowed purchase of workstation where some of the simulations are performed. SRH would like to acknowledge the financial support from the IOE endowed postdoctoral position at IISc, Bangalore, India (IOE-PDF/CHEP/80011423-1042 and the Indo-French Centre for the Promotion of Advanced Research (CEFIPRA) for support of the proposal 6704-4 under the Collaborative Scientific Research Programme.}
\appendix
\section{Derivation of the flow equations}\label{App1}
In this appendix we derives the flow equation for different parameters in the theory. Starting from the \weq, one can write the flow equation for the two point vertex function as
\beq\label{tga2}
\p_{t}\ga2\= \tr\left[ \p_{t}\rk\l \frac{(\Gamma^{(3)})^{2}}{\l \ga2+\rk\r^{3}}\r\right]-\frac{1}{2}\tr\left[ \p_{t}\rk\frac{\Gamma^{(4)}}{\l \ga2+\rk\r^{2}} \right].
\eeq
For the ansatz (Eq. \ref{ansatz}) we used for the $\gak$, LHS of the above quation is given as
\beq
\p_{t}\ga2\= \p_{t}Z_{1} p^{2}+\frac{\p_{t}\lambda^{2}}{p^{2}}+\p_{t}\uk''\l\phi\r.
\eeq
So, one can obtain $\p_{t}\z1$ and $\p_{t}\lam2$ by taking a $\p^{2}$ and $1/p^{2}$ coefficients of the RHS.  We only focus on the first term in the RHS of Eq. \ref{tga2} as in the same way one can show that the second term is independent of external momenta $p$, which is the same for local theory. Now,
\newpage
\beq
&\tr\left[ \p_{t}\rk \l \frac{(\Gamma^{(3)})^{2}}{\l \ga2+\rk\r^{3}}\r\right] = \tr \left[\sum_{q_{1}...q_{4}}G\ket{q_{1}}\bra{q_{1}}\g3\ket{q_{2}}\bra{q_{2}}G\ket{q_{3}}\bra{q_{3}}\g3\ket{q_{4}}\bra{q_{4}}G\p_{t}\rk\right]\\
\= \int_{q,q_{1}...q_{2}}G\l q,-q_{1}\r \g3\l p,q,-q_{2}\r G\l q_{2},-q_{3}\r\g3\l -p,q_{3},-q_{4}\r G\l q_{4},-q\r \p_{t}\rk\l q\r.
\eeq
Now using the momentum conservation of the propagators, in each vertex and noting that the $\g3 = \uk^{(3)}$ we can re write this as
\beq\label{z1eq1}
\tr\left[ \p_{t}\rk \l \frac{(\Gamma^{(3)})^{2}}{\l \ga2+\rk\r^{3}}\r\right] \=  \l \uk^{(3)}\r^{2}\int_{q}G^{2}_{q}G_{p+q}\p_{t}\rk,
\eeq
where 
\beq\label{nonlprop}
G\l q,-q\r = G_{q}\= \frac{1}{Z_{1}q^{2}+\frac{\lambda^{2}}{q^{2}}+\uk''+\rk} = \frac{\theta\l k^{2}-q^{2}\r}{f(k)}+\frac{\theta\l q^{2}-k^{2}\r}{f(q)},
\eeq
where $f(q)= \z1 q^{2}+\uk''+\lam2/q^{2}$. Note that all the $p$ dependence is coming from $G_{p+q}$ alone. However,  $G_{p+q}$ is known to be analytic in $p+q$ \cite{NARAIN2019143}, as one can write $G_{q}$ as the sum of propagators of two local theories as
\beq
G_{q}= \frac{A}{q^{2}+r_{-}^{2}}+\frac{B}{q^{2}+r_{+}^{2}}.
\eeq
So one can conclude that $\p_{t}\lam2 = 0$.  Inorder to obtain the $p^{2}$ coefficient we Taylor expand $G_{p+q}$ around $p=0$ as 
\beq
G_{p+q}\= G_{q}+p_{\mu}\frac{\p}{\p q_{\mu}}G_{q}+\frac{1}{2}p_{\mu}p_{\nu}\frac{\p^{2}}{\p q_{\mu}\p{q_{\nu}}}G_{q}+ O(p^{3}).
\eeq
However, in our model, $G_{q}$ is a function of $q^{2}$ so we can use the modified expansion in which
\beq
p_{\mu}p_{\nu}\= \frac{p^{2}}{d}\delta_{\mu\nu}.
\eeq 
Then by comparing the coefficients on the Eq. \ref{tga2}
\beq
\p_{t}\z1 =   \l \uk^{(3)}\r^{2}\int_{q}G^{2}_{q}\p_{t}\rk\l \frac{4 q^{2}}{2d}G''+G'\r,
\eeq
where $G'$ is the derivative w.r.t $q^{2}$. With our regulator choice (Eq. \ref{regulator}), one can show that $G'$ term won't contribute and we can do the remaining momentum integral  using the identity
\beq
\int_{x} \theta\l x-a\r\delta\l x-a\r = \frac{1}{2},
\eeq
we can obtain
\beq
\p_{t}\z1 = \mu_{d}k^{d+2}\l \uk^{(3)}\r^{2}\l\z1 -\frac{\lam2}{k^{4}}\r^{2}\l \z1 k^{2}+\frac{\lam2}{k^{2}}+\uk^{(2)}\r^{-4},
\eeq
where $\mu_{d}^{-1}= \l 4\pi\r^{d}\Gamma\l d/2+1\r$.  Now, directly from the \weq, using the leading order term in the LPA one can write
\beq
\p_{t}\uk \= \frac{1}{2}\int_{q}G_{q}\p_{t}\rk = -\mu_{d}k^{d+2}\l \z1 k^{2}+\frac{\lam2}{k^{2}}+\uk^{(2)}\r^{-1}\l \z1-\frac{\lam2}{k^{4}}+\frac{\p_{t}\z1}{d+2}\r.
\eeq
In our analysis we are considering the potential with non zero vacuum expectation value.  So we also need to consider the flow equation for the vacuum expectation value ($\kappa$). One can obtain this from the condition for minimum of the potential i.e.,
\beq
\frac{\p \uk}{\p \rho}\Big|_{\rho=\kappa} \= 0.
\eeq
Taking the total derivative over $t$ one can get
\beq\label{tkappa}
\p_{t}\kappa\= -\mu_{d}k^{d+2}\l \z1-\frac{\tlam}{k^{4}}+\frac{\p_{t}\z1}{d+2}\r\l 3\uk''+2\rho\uk'''\r\ll\uk''\l \z1 k^{1}+\frac{\lam2}{k^{2}}+\uk'+2\rho\uk''\r\rr^{-1}.
\eeq

\subsection*{Flow equation of wavefunction coefficient from local theory}\label{App12}
In this subsection we derives the flow equation for the $\z1$ from the local theory discussed in section \ref{seclocal}.  Consider
\beq
\p_{t}\gg2\=\begin{pmatrix}
\p_{t}\ga2_{11} & 0\\
0& 0
\end{pmatrix} .
\eeq
where we took  $\p_{t}\lambda=0$ and $\p_{t}Z_{2}=0$.  Taking the second derivative of Eq. \ref{loweq} with respect to $\phi$ gives
\beq
\p_{t}\Gamma^{(2)}_{11} \=  p^{2}\p_{t}\z1 +\p_{t}\mu_{1} =\frac{1}{2}\tr\left[ \p_{t}\rkk\l\frac{2\l\ggg\r^{2}}{\l \gg2+\rkk\r^{3}} -\frac{\mathbb{\Gamma}^{(4)}_{k}}{\l \gg2+\rkk\r^{2}}\r\right].
\eeq
Similar to the non local theory (See Appendix \ref{App1}), the seond term is independent of the external momentum. So in order to obtain $\p_{t}\z1$, we only need to consider the first term.  Then,
\beq
\p_{t}\Gamma^{(2)}_{11}\=\tr\begin{pmatrix}
G_{11}^{3}\l\Gamma^{(3)}\r^{2}\p_{t}R_{1} & G_{11}^{2}G_{12}\l \Gamma^{(3)}\r^{2}\p_{t}R_{2}\\
G_{11}^{2}G_{12}\l \Gamma^{(3)}\r^{2}\p_{t}R_{1}& G_{11}G_{12}G_{21}\l \Gamma^{(3)}\r^{2}\p_{t}R_{2}
\end{pmatrix} ,
\eeq
where 
\beq
\l \gg2+\rkk\r^{-1} \equiv \mathbb{G} = \begin{pmatrix}
G_{11} & G_{12}\\
G_{21}& G_{22}
\end{pmatrix} ,
\eeq
with
\beq\label{gii}
G_{11}\= \frac{\ga2_{22}+R_{2}}{\l\ga2_{11}+R_{1}\r\l\ga2_{22}+R_{2}\r-\lam2} =\frac{\theta\l k^{2}-q^{2}\r}{f(k)}+\frac{\theta\l q^{2}-k^{2}\r}{f(q)},\\
G_{12}=G_{21}\=\frac{\lambda}{\l\ga2_{11}+R_{1}\r\l\ga2_{22}+R_{2}\r-\lam2} = \frac{\lambda\theta\l k^{2}-q^{2}\r}{k^{2}f(k)}+\frac{\lambda\theta\l q^{2}-k^{2}\r}{q^{2}f(q)},\\
G_{22}\=  \frac{\ga2_{11}+R_{1}}{\l\ga2_{11}+R_{1}\r\l\ga2_{22}+R_{2}\r-\lam2},
\eeq
where we have used the explicit form of $R_{i}$ (See Eq. \ref{lolitim}) with $f(q)= \z1 q^{2}+\uk''-\lam2/q^{2}$  and
\beq
\ggg \= \begin{pmatrix}
\Gamma^{(3)} & 0\\
0& 0
\end{pmatrix} .
\eeq
Note that $G_{11}$ is equivalent to the non local propagator once we take $\lam2\rightarrow -\lam2$ (See Eq. \ref{nonlprop}). After taking the trace over the matrix index and introducing the momentum basis (See above section), we can write that
\beq\label{locz1}
\p_{t}\ga2_{11}\= \l \uk^{(3)}\r^{2}\int_{q}G_{11}(p+q)\l G_{11}^{2}(q) \p_{t}R_{1}+G_{12}^{2}(q) \p_{t}R_{2}\r.
\eeq
Now using the explicit for Eq. \ref{gii}, one can see that
\beq
G_{11}^{2}(q) \p_{t}R_{1}+G_{12}^{2}(q) \p_{t}R_{2} \= f^{-2}(k)\l \p_{t}\z1\l k^{2}-q^{2}\r+ 2 k^{2}\z1 +\frac{2\lam2}{k^{2}}\r.
\eeq
With this we can see that  Eq. \ref{z1eq1} is equivalent to  Eq. \ref{locz1}  when $\lam2\rightarrow -\lam2$.  Then following the same steps as in the non local theory, one can re derive the formula for $\p_{t}\z1$ from the local theory itself.
\vspace{-1em}
\section{Shooting from infinity}\label{appshoot}
In this appendix we solve the flow equation of the scaled potential  (Eq. \ref{nlscflow}) using the shooting from infinity method \cite{Bridle2014} in $d=3$ dimension and the analysis can be extended to other dimensions also.  First we look at the asymptotic solution of the differential equation for large field values, which one can write as
\beq\label{shootingasym}
 \lim_{\phi\rightarrow \infty}\tu\l \phi\r \sim A \phi^{\frac{6}{5}}-\frac{1}{3}+\frac{6 A}{125\phi^{\frac{4}{5}}}-\frac{36}{4375}\frac{A^{2}}{\phi^{\frac{8}{5}}}+ ...
\eeq
\begin{figure}[H]
\centering
\includegraphics[height=9cm, width=13cm]{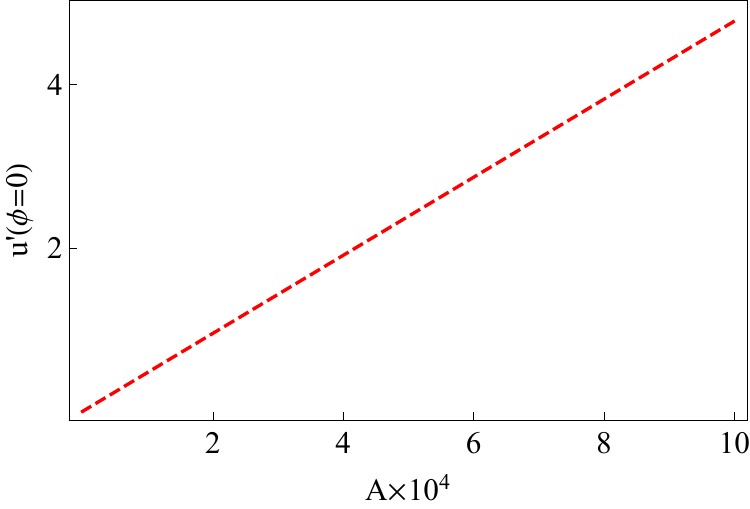}
\caption{Plot shows the $u'(\phi=0)$ as function of $A$. From the plot one can see that there is only one value ($A=0$), which satisfies the required condition i.e, $u'(\phi=0)=0$. }
\label{shooting}
\end{figure}
Compare to the standard local theory, where the leading part of the asymptotic solution goes as $\sim \phi^{6}$, our solution goes slowly to the infinity.  Now we numerically solve Eq. \ref{nlscflow} with the initial conditions calculated from Eq. \ref{shootingasym} for some large values of $\phi$ such that the series is convergent. As our asymptotic solution (Eq. \ref{shootingasym}) goes slowly to the infinity, we require to take large values of $\phi\sim 10^{6}$, which can be achieved by scaling the field value appropriately. With this one can get the solution of Eq. \ref{nlscflow} as a function of $A$.  Now we plot the solution for different values of $A$ and choose the ones which satisfy $u'(\phi=0)=0$ (see Fig. \ref{shooting}).  From Fig. \ref{shooting}, we can see that there is only one solution which satisfy $u'(\phi=0)=0$, which is $A=0$, which corresponds to the constant solution or the Gaussian fixed point.
\bibliographystyle{apsrev4-1}
\bibliography{reference}
\end{document}